\documentclass[1pts,showpacs,preprint,aps]{revtex4}
\usepackage{graphicx}
\usepackage{dcolumn}
\usepackage{bm}
\usepackage{amssymb}
\usepackage{amsmath}
\usepackage{mathrsfs}
\begin{document}
\setcounter{page}{1}
\vskip 0.5cm
\title
{The Vainshtein conditions: The Vainshtein mechanism in terms of St\"uckelberg functions}
\author
{Ivan Arraut}
\affiliation
{State Key Laboratory of Theoretical Physics, Institute of Theoretical Physics,
Chinese Academy of Science,
Beijing 100190, China}

\begin{abstract}
Here I develop the simplest method in order to evaluate whether or not the Vainshtein mechanism can operate for a given set of parameters in a given solution. The method is based on the formulation of the mechanism in terms of the St\"uckelberg functions given in Int.J.Mod.Phys. D24 (2015) 1550022 and arXiv:1305.0475 [gr-qc]. In such a case, the Vainshtein scale appears as an extremal condition of the dynamical metric. If we fix the graviton mass, we can define the effective Vainshtein scale. Then for parameters where the Vainshtein scale vanishes or becomes smaller than the gravitational radius, the mechanism should be absent. At the other extreme, if the Vainshtein scale is finite or infinite, then the mechanism can operate. For consistency, if we define the Vainshtein scale as an invariant, then we should expect the effective graviton mass to become very large when the Vainshtein mechanism operates. On the other hand, if the mass scale tends to zero, then the extra-degrees of freedom are free to propagate. For clarity, here I analyze the effective mass behavior for the different type of modes.  
\end{abstract}
\maketitle 

\section{Introduction}
The first massive gravity formulation was done at the linear level by Fierz and Pauli \cite{Fierz}. This first attempt failed after comparing its predictions with the solar system observations. The reason for the failure of the theory at this level is the additional attractive effect due to the coupling between the scalar component and the trace of the energy-momentum tensor. This coupling remains even at the mass-less limit and it is known as the vDVZ discontinuity \cite{vDVZ}. Then after Vainshtein proposed that it is possible to recover the predictions of GR at scales near the source if we consider a non-linear formulation of massive gravity instead of the standard linear version \cite{Vainshtein}. Introducing non-linearities however, makes the theory pathological due to the appearance of a ghost \cite{Deser}. In fact, the recovery of GR at scales of the solar system was due to the presence of this ghost since the negative kinetic energy term reproduced by the ghost exactly cancels the positive kinetic energy contribution coming from the scalar component. The details about this physical fact are explained in \cite{Deff}. Due to the presence of a ghost, the non-linear theory was considered pathological. Later de-Rham, Gabadadze and Tolley discovered that by tuning in an appropriate way the different parameters of the potential, it was then possible to reproduce a ghost-free formulation of massive gravity. The essential idea is that the pathological terms of the action, after being grouped with higher order terms, are total derivatives and they will never appear in the equations of motion. The theory formulated in this way is called dRGT \cite{derham}. Inside this theory, the recovery of GR at scales of the solar system is due to the non-linearities, relevant below the Vainshtein scale which is approximately $r_V\backsim (GM/m^2)^{1/3}$. In \cite{My papers}, the author formulated the Vainshtein mechanism in terms of St\"uckelberg functions. The Vainshtein scale then appeared as an extremal condition of the dynamical metric in unitary gauge. This is equivalent to say that the scale is an extremal condition of the massive action. In this manuscript, I use this formulation in order to explain for which set of parameters it is possible to expect the Vainshtein mechanism to operate and for which set of parameters, the mechanism is absent. The theory has generically three free-parameters, namely, the graviton mass, and the two free-parameters of the potential. By imposing the Schwarzschild de-Sitter background condition, the number of free-parameters is reduced to two. If we fix the graviton mass parameter, then we can define an effective Vainshtein scale. For the set of parameters where this scale vanishes or becomes smaller than the gravitational radius $GM$, the mechanism must be absent. On the other hand, for the set of parameters where the mechanism appears, the effective Vainshtein scale must be finite. If it becomes infinite, then the screening mechanism should appear at any scale larger than $GM$. An alternative point of view is to set the Vainshtein scale as an invariant. Then we can define an effective graviton mass. At scales where the effective graviton mass almost vanishes, the mechanism should be expected to be absent and the massive gravitons can propagate freely everywhere. On the other hand, if the effective graviton mass is finite, the mechanism must appear. If the effective graviton mass is infinite, then the massive gravitons cannot propagate so far and then GR can in principle be recovered. In order to prove the consistency of these previous statements, I analyze the effective graviton mass for the different modes separately, namely, scalar, vectorial and the purely radial component in a free-falling frame of reference. In such a case, the analysis is more transparent and it also takes into account the role of the St\"uckelberg function when the family of solutions with one free-parameter is analyzed \cite{Kodama}.                

\section{The Schwarzschild de-Sitter solution in dRGT}   \label{thisla}   

The black-hole solutions inside the non-linear formulation were already found in \cite{Gaba, Koyama}. The simplest and most generic solution for the spherically symmetric solution, was found in \cite{Kodama}. All the solutions satisfying the stationary condition, are defined as:

\begin{equation}
ds^2=G_{tt}dt^2+G_{rr}S^2dr^2+G_{rt}(drdt+dtdr)+S^2r^2d\Omega_2^2,
\end{equation}
where:

\begin{eqnarray}   \label{eq:drgt metric}
G_{tt}=-f(Sr)(\partial_tT_0(r,t))^2,\;\;\;\;\;\;G_{rr}=-f(Sr)(\partial_rT_0(r,t))^2
+\frac{1}{f(Sr)},\;\;\;\;\;\;\nonumber\\G_{tr}=-f(Sr)\partial_tT_0(r,t)\partial_rT_0(r,t),
\end{eqnarray}
and $f(Sr)=1-\frac{2GM}{Sr}-\frac{1}{3}\Lambda (Sr)^2$, with $S$ being the scale factor which depends on the free-parameters of the theory. In this previous solution, all the degrees of freedom are inside the dynamical metric. Here we are in unitary gauge and then the fiducial metric is:

\begin{equation}   \label{eq:drgt metric223}
f_{\mu\nu}dx^\mu dx^\nu=-dt^2+dr^2+r^2(d\theta^2+r^2sin^2\theta).
\end{equation}
The solution (\ref{eq:drgt metric}) is equivalent to:

\begin{equation}   \label{eq:drgt metric2}
ds^2=-f(Sr)dT_0(r,t)^2+\frac{S^2dr^2}{f(Sr)}+S^2r^2d\Omega^2,
\end{equation}
where $T_0(r,t)$ corresponds to the St\"uckelberg function. The non-triviality of $T_0(r,t)$ contains the information of the extra-degrees of freedom \cite{My papers, Kodama, K, On the app}. In fact, it has been demonstrated in \cite{newlaaleji}, that $T_0(r,t)$ represents a preferred time-direction of the theory when the extra-degrees of freedom become relevant. The solution (\ref{eq:drgt metric2}) can be classified in two families.

\subsubsection{Family 1: One free-parameter and the St\"uckelberg function arbitrary}
For this family of solutions, the St\"uckelberg function is also a parameter. In other words, there is no rule or constraint telling us how this function should be specified. Then in reality we still have two free-parameters, namely, one inside the potential and another represented by $T_0(r,t)$, which has a non-trivial relation with the graviton mass. For this case, the relation between the cosmological constant $\Lambda$ and the graviton mass, is given by \cite{Kodama}:

\begin{equation}    \label{eq:cosmolala}   
\Lambda=\frac{m^2}{\alpha}, 
\end{equation}
where $\alpha$ corresponds to the free-parameter of the massive action in agreement with the notation given in \cite{Kodama}. The relation between the parameters of the massive action is defined in agreement with the condition $\beta=\alpha^2$. The result (\ref{eq:cosmolala}) is telling us that the graviton mass is another free-parameter and its dependence with respect to $\alpha$ must take the form $m^2\backsim \alpha$ if we want to keep $\Lambda$ as a fundamental constant. Here however, I want to keep the Vainshtein scale (not the cosmological constant) as an invariant quantity. In agreement with the results obtained in \cite{newlaaleji}, the equivalence principle is not satisfied for this family of solutions due to the preferred notion of time in agreement with the function $T_0(r,t)$. The equivalence principle is only satisfied if $T_0'(r,t)=0$ and this is equivalent to the absence of the extra-degrees of freedom for this family of solutions.     

\subsubsection{Family 2: Two free-parameters and the St\"uckelberg function constrained}

For the family of solutions with two free-parameters and the St\"uckelberg function constrained, the relation between the cosmological constant and the graviton mass is given by \cite{Kodama}:

\begin{equation}    \label{eq:cosmo}   
\Lambda=-m^2\left(1-\frac{1}{S}\right)\left(2+\alpha-\frac{\alpha}{S}\right).
\end{equation}
Depending on the interpretation of the theory, the cosmological constant can be zero if we fix the graviton mass as an invariant. If however, what we keep as an invariant is the cosmological constant, then the value taken by the graviton mass might change. In particular, the relation $\beta=(3/4)\alpha^2$ can give us a zero $\Lambda$ value or an infinite graviton mass, depending on the point of view. Independent of the interpretation used, the St\"uckelberg function has to satisfy the constraint \cite{Kodama}:

\begin{equation}
(T_0'(r,t))^2=\frac{1-f(Sr)}{f(Sr)}\left(\frac{S^2}{f(Sr)}-\dot{T}_0^2\right).
\end{equation}
A global solution of this previous constraint, is given by the Finkelstein-type form \cite{Kodama}:

\begin{equation}   \label{eq:Tzero}
T_0(r,t)=St\pm\int^{Sr}\left(\frac{1}{f(u)}-1\right)du.
\end{equation}
For the family of solutions with one free-parameter but with the St\"uckelberg function arbitrary, in principle there is no clear solution for $T_0(r,t)$. The arbitrariness of $T_0(r,t)$ was considered as pathological for the perturbative analysis considered in \cite{Kodama}. An alternative interpretation, was however considered in \cite{On the app}. The connection of this arbitrariness with the Higgs mechanism at the graviton level was analyzed in \cite{newlaaleji}. Note that interestingly, for the family of solutions analyzed in this section, the equivalence principle is recovered. In fact, from eq. (\ref{eq:Tzero}), it is easy to demonstrate that $T_0'(r,t)=0$ if $f(Sr)\to 1$. Then, for the first family of solutions (one free-paramater), we can conclude that the equivalence principle is in reality hidden. In other words, the principle which is absent from the family of solutions with one free-parameter and $T_0(r,t)$ arbitrary, is recovered when we select as free-parameters those inside the massive action \cite{newlaaleji}. Note that the St\"uckelberg function depends on the graviton mass through the function $f(Sr)$ defined previously. Then the arbitrariness on $T_0(r,t)$ defined for the previous family of solutions, is equivalent to an arbitrariness on the graviton mass value. Then for the first family of solutions, the two free-parameters are the graviton mass and $\alpha$.     
\section{The Vainshtein conditions}   

The Vainshtein conditions were derived originally by the author in \cite{My papers}. They express the Vainshtein scale as an extremal condition of the dynamical metric. This result helps us to find a relation between the St\"uckelberg function $T_0(r,t)$ and the function $f(Sr)$ contained inside the components of the dynamical metric. The key relations are:

\begin{eqnarray}   \label{eq:thisoneman}
\partial_rT_0(r,t)=0\to r<<r_V, \;\;\;\;\;\;\;\;\partial_r T_0(r,t)\neq0\to r>>r_V\nonumber\\
T_0''(r,t)=0\to r=r_V,
\end{eqnarray}
with $r_V$ being the Vainshtein scale. The vanishing condition $T_0'(r,t)=0$, guarantees the recovery of GR in the usual Schwarzschild-like coordinates. The same set of relations, could also be derived from the most direct condition $dU(g, \phi)=0$ \cite{My papers}. The result is independent of the exact form of the potential $U(g, \phi)$.   
\section{The Vainshtein conditions: Extremal conditions on the massive action}   \label{eq:JustVainshtein}

The Vainshtein scale, corresponding to an extremal conditions of the dynamical metric, can also be derived directly from the potential $U(g,\phi)$ as has been explained in the previous section. The key point is to understand that in unitary gauge, namely, when all the degrees of freedom (Nambu-Goldstone bosons) are inside the dynamical metric (eaten up by the dynamical metric), then any variation of the potential, is directly related to the variations on the dynamical metric. The relevant  mathematical condition is \cite{My papers}:

\begin{equation}   \label{eq:ptm}
dU(g,\phi)=\left(\frac{\partial U(g, \phi)}{\partial g}\right)_\phi dg+\left(\frac{\partial U(g, \phi)}{\partial g}\right)_g d\phi.
\end{equation}
The importance of the Vainshtein conditions is that the details of the potential are not relevant at all in order to derive the Vainshtein scale. All the information about the Vainshtein scale is contained inside the dynamical metric if it contains all the degrees of freedom. For stationary cases, where the metric components are time-independent, then eq. (\ref{eq:ptm}) becomes:

\begin{equation}   \label{eq:ptm2}
dU(G)=\left(\frac{\partial U(G)}{\partial g}\right)dg=0,
\end{equation}
and this is just equivalent to:

\begin{equation}   \label{eq:this one}
dg=0,
\end{equation}
if the matrix $\left(\frac{\partial U(G)}{\partial g}\right)$ is non-singular. The condition (\ref{eq:this one}), has to be satisfied by each component of the metric containing the information of the extra-degrees of freedom. The conditions are valid at the Vainshtein scale. Then whenever we want to calculate the Vainshtein radius of the theory, all what we have to do is to translate all the degrees of freedom to the dynamical metric and then apply the conditions (\ref{eq:this one}) and finally solve algebraically for $r=r_V$ \cite{My papers}.
\section{The parameter-dependent Vainshtein scale: The case with one free-parameter with a fixed graviton mass}   \label{eq:mfeva}   

Here I derive the effective Vainshtein scale for the two type of solutions obtained in \cite{Kodama}. The first solution to be considered is the one with one free-parameter but with the St\"uckelberg function arbitrary. In this case, in principle there is no correspondence between $T_0(r,t)$ and the metric function $f(Sr)$. The dynamical metric in unitary gauge takes the explicit form given by eq. (\ref{eq:drgt metric}) but taking into account that $T_0(r,t)$ is in principle arbitrary. Then although the location of the Vainshtein scale in terms of the St\"uckelberg function is not clear at all, we can assume that the Vainshtein conditions can still be applied to the dynamical metric, providing then the information about the location of the scale after which the effects of accelerated expansion have to be taken into account. For the stationary case, in agreement with \cite{My papers}, we can define the effective Vainshtein scale as:

\begin{equation}   \label{eq:this oneis not}
f'(Sr)=0, \;\;\;\;\;\;\;\;\to\;\;\;\;\;\; r=r_V,
\end{equation}
\begin{equation}   \label{eq:this oneis neither}
r_V=\frac{1}{S}\left(\frac{3GM}{\Lambda}\right)^{1/3}=\frac{\alpha+1}{\alpha^{2/3}}\left(\frac{3GM}{m^2}\right)^{1/3}.
\end{equation}
This is an effective value of the Vainshtein radius depending on the free-parameters of the theory. In this case however, the possible effects of the St\"uckelberg function $(T_0(r,t))$ have not been taken into account. Later in this manuscript I will analyze the role of $T_0(r,t)$ but inside the definition of effective graviton mass. The analysis done in Sec. (\ref{eq:Perturbative}), takes into account the effective mass for each mode if we keep the Vainshtein scale as an invariant.   
\begin{figure}
	\centering
		\includegraphics[width=0.6\textwidth, natwidth=800, natheight=400]{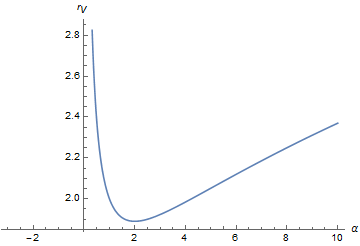}
	\caption{The effective Vainshtein scale for a fixed graviton mass. Note that the (effective) Vainshtein scale becomes asymptotically infinite when $\alpha\to0$ and $\alpha\to\infty$. The Vainshtein scale has a minimum for $\alpha=2$. The factor $\left(\frac{3GM}{m^2}\right)^{1/3}$ has been normalized to be one. This figure also shows the qualitative behavior of the Vainshtein scale for one of the branches of the solution with $\beta=0$ and $\alpha$ arbitrary.}
	\label{fig:momoko3}
\end{figure}
Fig. (\ref{fig:momoko3}) shows the behavior of the scale (\ref{eq:this oneis neither}) as a function of $\alpha$. Note that the Vainshtein scale is non-zero for any positive value taken by the parameter $\alpha$. This means that the Vainshtein mechanism will always appear for this solution. When $r_V\to\infty$, GR should be recovered everywhere. However, if we analyze term by term in the function $f(Sr)$, given in terms of $\alpha$ by:

\begin{equation}   \label{eq:this onetoo}
f(Sr)=1-\frac{2GM(\alpha+1)}{\alpha r}-\frac{1}{3}\frac{\alpha^2r^2}{(\alpha+1)^2},
\end{equation}
then it is clear that GR is recovered everywhere for $\alpha\to\infty$. However, for $\alpha\to0$, the Newtonian term $\frac{2GM(\alpha+1)}{\alpha r}$ becomes dominant, rescaling the effective Newtonian constant and making gravity very strong and attractive. Then the effective Schwarzschild radius (gravitational radius) becomes very large. This means that even if the Vainshtein mechanism operates everywhere outside the gravitational radius, this part of the solution is unphysical from the experimental point of view. However, the main point here is that as far as the Vainshtein scale is non-zero, the mechanism is able to operate outside the gravitational radius. The case explored here is equivalent to the one with the condition $\beta=\alpha^2$.      	
\subsection{The case with two free-parameters and the St\"uckelberg function constrained and a fixed graviton mass}

\begin{figure}
	\centering
		\includegraphics[width=0.6\textwidth, natwidth=560, natheight=350]{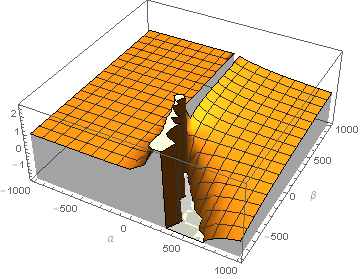}
	\caption{The behavior of the scale factor $S$ as a function of the free-parameters of the theory. This plot corresponds to the negative root square branch in agreement with eq. (\ref{eq:this onetoooo}). This picture is the mirror image of the positive root square branch of $S$.}
	\label{fig:momoko3333557778899}
\end{figure}

For this case, there is a direct correspondence between $T_0(r,t)$ and $f(Sr)$. However, still we can define the Vainshtein scale in the same way as in the previous case. It is given by the result (\ref{eq:this oneis neither}), but with the cosmological constant now defined as in eq. (\ref{eq:cosmo}). The scale factor $S$ is given by \cite{Kodama}:

\begin{equation}   \label{eq:this onetoooo}
S_{1,2}=\frac{\alpha+\beta\pm\sqrt{\alpha^2-\beta}}{1+2\alpha+\beta}.
\end{equation}
The solution can then be classified in terms of the different values taken by the parameters $\alpha$ and $\beta$. Figure (\ref{fig:momoko3333557778899}) shows the behavior of $S$ as a function of the two free-parameters of the theory for the negative root square branch. The same factor but for the positive branch is just the mirror image of this figure. In what follows, I will analyze two cases.   
\subsection{Case i). $\beta=0$, $\alpha$ arbitrary and a fixed graviton mass}

For this case, there are two possible values for the scale factor $S$. They are:

\begin{equation}   \label{eq:this onetoooodudu}
S_1=\frac{2\alpha}{1+2\alpha},\;\;\;\;\;\;\;\;\;\;\;S_2=0.
\end{equation}
The cosmological constant as a function of the graviton mass is in agreement with eq. (\ref{eq:cosmo}) the following:

\begin{equation}   \label{eq:this onetoooo55}
\Lambda_1=-\frac{m^2}{S^2_1}(S_1-1)(2S_1+\alpha S_1-\alpha),\;\;\;\;\;\;\;\;\;\;\;\Lambda_2=-\frac{m^2\alpha}{S_2^2}.
\end{equation}
Note that $\Lambda_2$ diverges if $\alpha\neq0$. If $\alpha=0$, then we recover the special case of the previous section, namely, with one free-parameter but with the St\"uckelberg function arbitrary. This case is also divergent for $\alpha\to0$. Here we can define the parameter-dependent Vainshtein scale as the extremal condition of the dynamical metric in unitary gauge \cite{My papers}. The result is:

\begin{equation}   \label{eq:this onetoooo555}
r_{V1}=\frac{2\alpha+1}{(6\alpha^2)^{1/3}}\left(\frac{3GM}{m^2}\right)^{1/3},\;\;\;\;\;\;\;\;\;\;r_{V2}=-\left(\frac{3GM}{S_2m^2\alpha}\right)^{1/3}.
\end{equation}
It is evident from the result (\ref{eq:this onetoooodudu}) that $r_{V2}$ will diverge for any value of $\alpha$. This means that the screening effects are expected to appear everywhere up to the gravitational radius scale in this particular situation. However, it is easy to find that the effective gravitational radius is also divergent, then the Vainshtein mechanism and any physically relevant solution related to this branch is absent. For the case of $r_{V1}$, the behavior is just the same as in the case studied previously when $\beta=\alpha^2$. Fig. (\ref{fig:momoko3}) still represents the behavior for this case in the qualitative sense. In fact, whenever we fix the parameter $\beta$, independent on its relation with $\alpha$, the qualitative behavior of the effective Vainshtein scale will be the same.

\subsection{Case ii). Both, $\beta$ and $\alpha$ arbitrary and a fixed graviton mass}
 
\begin{figure}
	\centering
		\includegraphics[width=0.6\textwidth, natwidth=560, natheight=350]{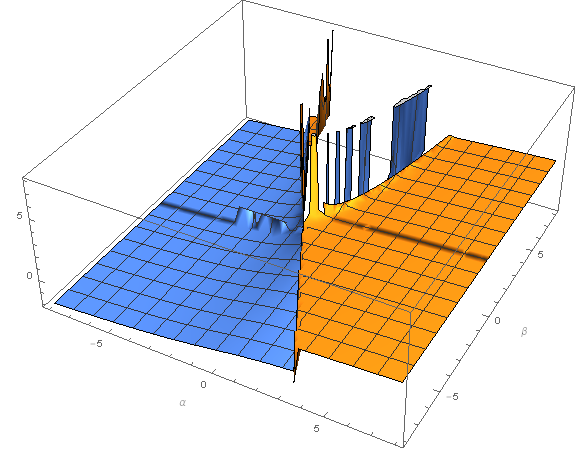}
	\caption{The parameter-dependent Vainshtein scale for a fixed graviton mass and arbitrary parameter combination. The peaks of the figure represent the largest possible Vainshtein scales. There is a region of discontinuity around the peaks representing the absence of the mechanism for the relevant parameter combination. The factor $\left(\frac{3GM}{m^2}\right)^{1/3}$ has been normalized to one. The yellow plot corresponds to the positive root square branch in agreement with eq. (\ref{eq:this onetoooo}) and the blue graphic corresponds to the negative root square.}
	\label{fig:momoko3333}
\end{figure}

\begin{figure}
	\centering
		\includegraphics[width=0.6\textwidth, natwidth=560, natheight=350]{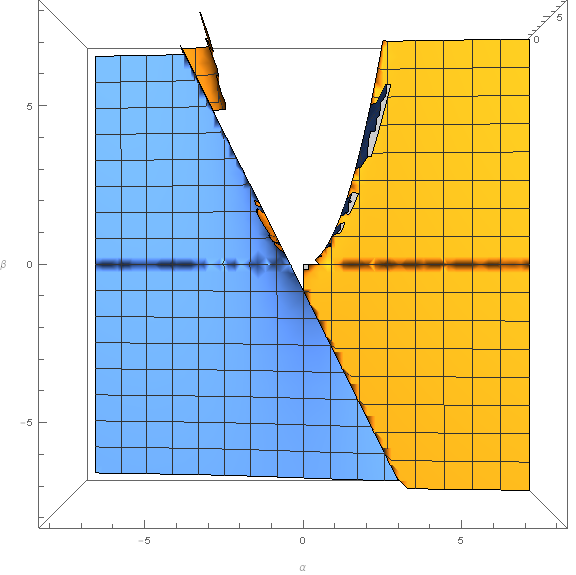}
	\caption{The same Figure (\ref{fig:momoko3333}), but from a different perspective. From this image the plane $\beta$-$\alpha$ can be appreciated.}
	\label{fig:momoko333355}
\end{figure}

\begin{figure}
	\centering
		\includegraphics[width=0.6\textwidth, natwidth=560, natheight=350]{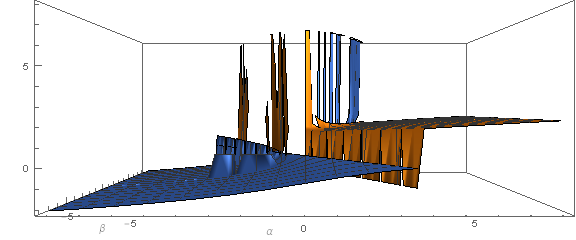}
	\caption{The same Figure (\ref{fig:momoko3333}), but from a different perspective.}
	\label{fig:momoko33335577788}
\end{figure}

For the two branches of solutions defined in agreement with eq. (\ref{eq:this onetoooo}), the parameter-dependent Vainshtein scale for this case becomes:
\begin{eqnarray}   \label{eq:this onetoooo555porque}
r_{V}=\left(\frac{1}{\alpha\pm\sqrt{\alpha^2-\beta}+\beta}\right)\left(\frac{3GM(1+2\alpha+\beta)^3\beta^2}{m^2(-2\alpha^3\pm2\alpha^2\sqrt{\alpha^2-\beta}+3\alpha\beta\mp2\beta\sqrt{\alpha^2-\beta})}\right)^{1/3}.
\end{eqnarray}
The two branches of solutions can be observed in Figs. (\ref{fig:momoko3333}), (\ref{fig:momoko333355}) and (\ref{fig:momoko33335577788}). Note that the effective Vainshtein scale has a discontinuity region. In agreement with the plots, the Vainshtein mechanism operates whenever we have a finite value for $r_V$.  

\section{The parameter-dependent graviton mass: The Vainshtein radius as an invariant}   \label{eq.Ivan}

\subsection{Family 1: One free-parameter and the St\"uckelberg function arbitrary}

\begin{figure}
	\centering
		\includegraphics[width=0.6\textwidth, natwidth=560, natheight=350]{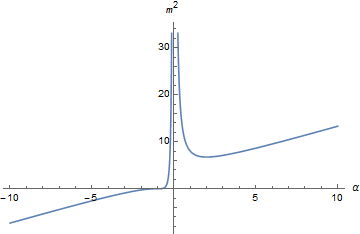}
	\caption{The parameter-dependent graviton mass as a function of the free-parameter of the theory. In this case, the Vainshtein scale is kept as an invariant. This case corresponds to the condition $\alpha=\beta^2$}
	\label{fig:momokola33335577788}
\end{figure}

If we fix as an invariant the Vainshtein scale, then we can construct an effective graviton mass. In this situation, we can interpret any of the two free-parameters, namely $\alpha$ or $\beta$ as the graviton mass. Here I will analyze the cases exposed in the previous sections but by keeping the Vainshtein scale as an invariant and then observing the behavior of $m^2$. For the case with one free-parameter corresponding to the Vainshtein scale given in eq. (\ref{eq:this oneis neither}), the effective graviton mass should behave as:

\begin{equation}   \label{eq:this onetoooo555miau}
m^2\backsim \frac{(\alpha+1)^3}{\alpha^2}.
\end{equation}  
Fig. (\ref{fig:momokola33335577788}) shows the behavior of $m^2$ for the different values taken by the free-parameter of the theory. Note that the effective graviton mass goes to infinity near $\alpha=0$. In the neighborhood of this value, the massive gravitons cannot propagate and then GR can in principle be recovered. This is consistent with the ideas expressed in the previous section suggesting that for a fixed graviton mass, infinite values of the effective Vainshtein scale correspond the range of parameters where GR can in principle be recovered. Note that in this case we are not taking into account the role of the St\"uckelberg function. This situation will be analyzed later in this manuscript.   

\begin{figure}
	\centering
		\includegraphics[width=0.6\textwidth, natwidth=560, natheight=350]{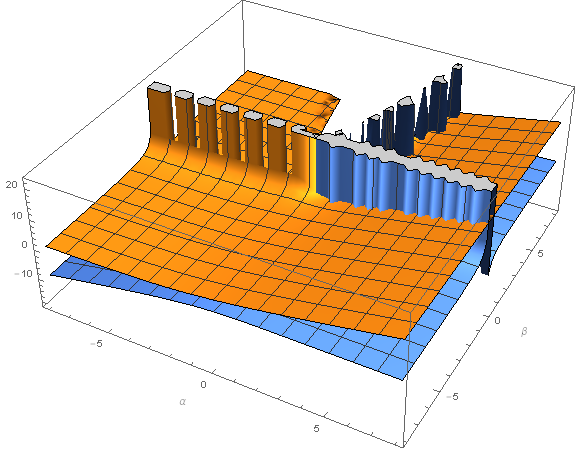}
	\caption{The parameter-dependent graviton mass as a function of the two free-parameters of the theory. In this case, the Vainshtein scale is kept as an invariant. The yellow plot corresponds to the positive brach of $S$ and the blue color to the negative one.}
	\label{fig:momokola33335577788Alejo}
\end{figure}
\begin{figure}
	\centering
		\includegraphics[width=0.6\textwidth, natwidth=560, natheight=350]{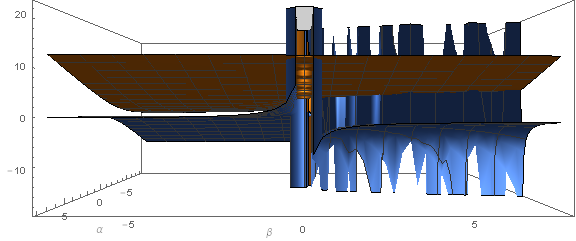}
	\caption{The parameter-dependent graviton mass as a function of the two free-parameters of the theory. This plot corresponds to the Fig. (\ref{fig:momokola33335577788Alejo}), but observed from a different perspective.}
	\label{fig:momokola33335577788Alejolala}
\end{figure}

\begin{figure}
	\centering
		\includegraphics[width=0.6\textwidth, natwidth=560, natheight=350]{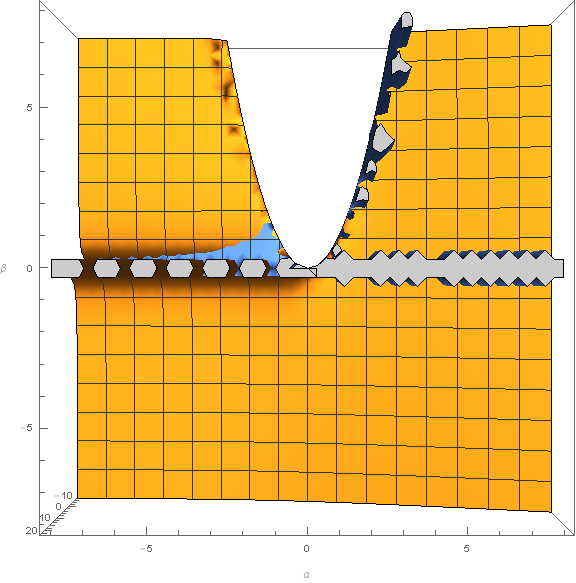}
	\caption{The parameter-dependent graviton mass as a function of the two free-parameters of the theory. This plot corresponds to the Fig. (\ref{fig:momokola33335577788Alejo}), but observed from above.}
	\label{fig:momokola33335577788Alejolalamiau}
\end{figure}
\subsection{Family 2: The parameter-dependent graviton mass for $\beta=0$, arbitrary $\alpha$ and a fixed Vainshtein scale}

If we fix the Vainshtein scale as an invariant, the parameter-dependent graviton masses corresponding to the case with $\beta=0$ and $\alpha$ arbitrary are in agreement with eqns. (\ref{eq:this onetoooo555}):

\begin{equation}   \label{eq:thismamamia}
m^2_1\backsim \frac{(2\alpha+1)^3}{\alpha^2}, \;\;\;\;\;\;\;\;\;\;\;\;\;\;\;\; m^2_2\backsim -\frac{1}{S_2\alpha}.
\end{equation}
For $m_2$, $S_2\to0$, which indicates that the parameter-dependent graviton mass diverges in that case for any value of $\alpha$. This again means that the massive gravitons cannot propagate for this branch of solutions and then GR could in principle be recovered everywhere. However, as has been discussed before, the gravitational radius for this solution also diverges, making it unphysical. The behavior of $m_1$ is exactly the same as in the case indicated in eq. (\ref{eq:this onetoooo555miau}) and then Fig. (\ref{fig:momokola33335577788}) is also appropriate for representing this branch of solutions. This is expected because whenever we fix the parameter $\beta$, the behavior of the solution with respect to $\alpha$ will not change in the qualitative sense. 

\subsection{Family 2: The parameter-dependent graviton mass for arbitrary $\beta$ and $\alpha$ parameters with a fixed Vainshtein scale}

In this case, the behavior of the parameter-dependent graviton mass is more complicated, but it can be visualized in Figs. (\ref{fig:momokola33335577788Alejo}), (\ref{fig:momokola33335577788Alejolala}) and (\ref{fig:momokola33335577788Alejolalamiau}). The regions where the parameter-dependent graviton mass takes large values, correspond to regions where in principle the massive gravitons can travel shorter distances. This is equivalent to an screening effect. Whenever the effective graviton mass is almost zero (but still finite), then the massive gravitons are able to travel larger scales and the screening effects are absent. However, although the method proposed here is simple, it does not take into account the possibility of having different behavior for different modes, namely, scalar, vector and tensor. This issue will be analyzed later in this manuscript after taking into account the perturbations of the action in a free-falling frame of reference.    

\section{Perturbative analysis}   \label{eq:Perturbative}    

In the previous sections, I worked around the background solutions without any concern about the perturbations. From the physical point of view however, if we are worried about the propagation of gravitons around some specific background, then we should analyze the action at the perturbative level. This provides a cleaner physical analysis and in this situation it is possible to understand the role of the St\"uckelberg function. The analysis at the background level, does not provide the opportunity for understanding the role of this function in the propagation of the graviton field. It has been demonstrated recently by the author that $T_0(r,t)$ can be considered as a preferred time-direction, breaking potentially the Lorentz symmetry for a free-falling observer \cite{newlaaleji}. In this section, I consider the perturbations of the action in agreement with a free-falling frame of reference. The method was developed in \cite{newlaaleji}, based in the generic perturbation analysis done in \cite{Kodama}. The purpose here is to find the effective mass for each of the modes, namely, scalar, vectorial and radial one. The most important mode is the scalar one because at the decoupling limit it generates the extra-attractive effect if the Vainshtein mechanism is absent. In general, the different modes of the graviton field, have different values of mass. 

\subsubsection{Family 1: One free-parameter and the St\"uckelberg function arbitrary}
 
This case was considered the most interesting one in \cite{newlaaleji}, when the spatial derivative of the St\"uckelberg function is non-vanishing. In this case, as has been explained in Sec. (\ref{thisla}), the equivalence principle is violated due to the preferred time direction which is a physical effect of the extra-degrees of freedom. The effect can also be observed as a spatial-dependence of the function $T_0(r,t)$. Under the assumption of stationary solutions, the potential expanded up to second order is given by: 

\begin{eqnarray}   \label{eq:acalanojo2miau}
\sqrt{-g}U(g,\phi)\approx \left(1+\frac{1}{2}h-\frac{1}{4}h^\alpha_{\;\;\beta}h^\beta_{\;\;\alpha}+\frac{1}{8}h^2\right)\left(\frac{2+6\alpha(1+\alpha)}{(1+\alpha)^4}\right)\nonumber\\
-\left(1+\frac{1}{2}h\right)\left(-\frac{h}{(1+\alpha)}+\frac{2T_0'(r,t)(1+\alpha)^2}{\alpha^3}h_{0r}-\frac{T_0'(r,t)^2(1+\alpha)^3}{\alpha^4}\right)+...     
\end{eqnarray}	
The matrix mass is obtained after finding the second derivative of the previous potential. It is given explicitly by:		

\begin{eqnarray}   \label{eq:acalanojo2miau22}
m^{\mu\nu}m^{\alpha\beta}=\left(-\frac{1}{2}\bar{\eta}^{\alpha\mu}\bar{\eta}^{\nu\beta}+\frac{1}{4}\bar{\eta}^{\mu\nu}\bar{\eta}^{\alpha\beta}\right)\left(\frac{2+6\alpha(1+\alpha)}{(1+\alpha)^4}\right)+\bar{\eta}^{\mu\nu}\bar{\eta}^{\alpha\beta}\left(\frac{1}{1+\alpha}\right)\nonumber\\
-\left(\bar{\eta}^{\mu\nu}\delta^\beta_{\;\;r}\delta^\alpha_{\;\;0}+\bar{\eta}^{\alpha\beta}\delta^\mu_{\;\;0}\delta^\nu_{\;\;r}\right)\frac{T_0'(r,t)(1+\alpha)^2}{\alpha^3},     
\end{eqnarray}		
where $\bar{\eta}^{\mu\nu}$ corresponds to the inverse of the dynamical metric in a free-falling frame of reference as has been defined in \cite{newlaaleji}. It is defined as:

\begin{equation}   \label{eq:acalanojo234}
ds^2=S^2\left(-dt^2+dr^2\left[1-\left(\frac{T_0'(r,t)}{S}\right)^2\right]-2\frac{T_0'(r,t)}{S}dtdr+r^2d\Omega^2\right).
\end{equation}
Without solving for the eigenvalues, we can understand the behavior for the graviton mass modes. The matrix (\ref{eq:acalanojo2miau22}) can in principle be diagonalized. However, it is not necessary at the moment of analyzing the different modes. We can find the value of the graviton mass for the different modes from eq. (\ref{eq:acalanojo2miau22}) if we download the indices by using the background metric. The general expression is given by:

\begin{equation}   \label{eq:acalanojosila}
m_{\mu\nu}=\bar{\eta}_{\gamma\mu}\bar{\eta}_{\beta\nu}m^{\gamma\beta}.
\end{equation}
Then the effective masses for the different modes are:

\begin{equation}
m_{00}=S^4m^{00}+S^2T_0'(r,t)^2m^{rr}+S^3T_0'(r,t)\left(m^{0r}+m^{r0}\right),
\end{equation}

\begin{equation}
m_{0r}=S^3T_0'(r,t)m^{00}-ST_0'(r,t)\left(S^2-T_0'(r,t)^2\right)m^{rr}+S^2T_0'(r,t)^2m^{r0}-S^2\left(S^2-T_0'(r,t)^2\right)m^{0r},
\end{equation}

\begin{equation}
m_{rr}=S^2T_0'(r,t)^2m^{00}+\left(S^2-T_0'(r,t)^2\right)^2m^{rr}-ST_0'(r,t)\left(S^2-T_0'(r,t)^2\right)\left(m^{0r}+m^{0r}\right).
\end{equation}
\begin{figure}
	\centering
		\includegraphics[width=0.6\textwidth, natwidth=560, natheight=350]{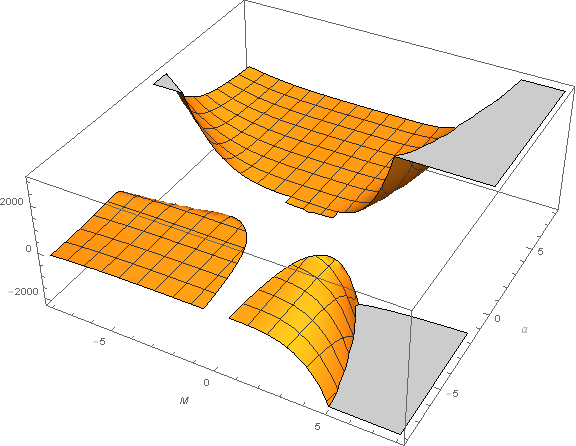}
	\caption{The square of the effective mass for the scalar mode. It can be observed that for a vanishing value of $T_0'(r,t)$ ($M$ in the figure), the scalar component of the graviton mass vanishes. Different branches appear for this solution.}
	\label{fig:sila}
\end{figure} 

\begin{figure}
	\centering
		\includegraphics[width=0.6\textwidth, natwidth=560, natheight=350]{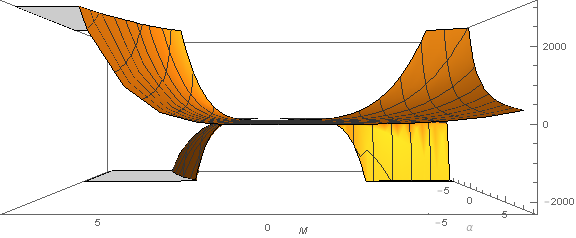}
	\caption{The same Fig. (\ref{fig:sila}) but observed from behind. Here it is clear that the existence scalar mode mass parameter vanishes when $T_0'(r,t)$ vanishes as a parameter. $T_0'(r,t)$ is represented by $M$ in the figure.}
	\label{fig:sila2}
\end{figure} 
 
\begin{figure}
	\centering
		\includegraphics[width=0.6\textwidth, natwidth=560, natheight=350]{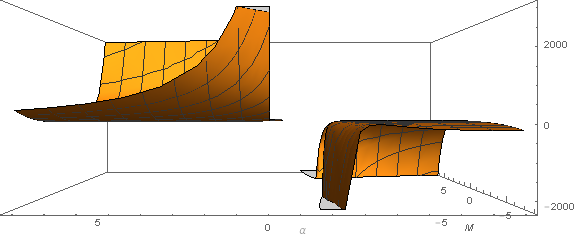}
	\caption{The same Fig. (\ref{fig:sila}) but observed from the left. From this angle, we can observe the scalar component of the parameter-dependent graviton mass as a function of $\alpha$. From the graphic, it is clear that the effective mass for this mode can become very large when $\alpha\to0$.}
	\label{fig:sila22}
\end{figure} 

\begin{figure}
	\centering
		\includegraphics[width=0.6\textwidth, natwidth=560, natheight=350]{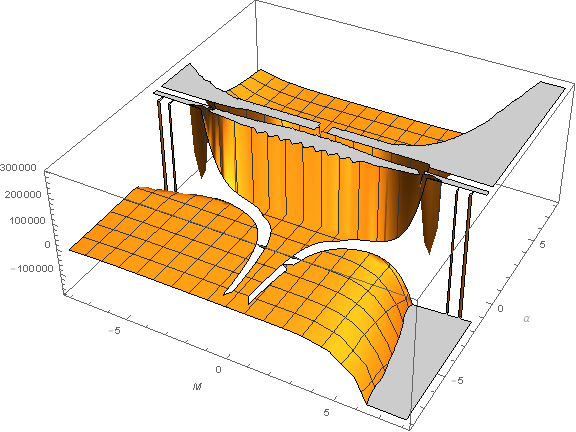}
	\caption{The parameter-dependent square of the graviton mass for the vectorial component. In this case, some divergences appear when $\alpha\to0$ independent on the values taken by $T_0'(r,t)$.}
	\label{fig:sila222}
\end{figure}

\begin{figure}
	\centering
		\includegraphics[width=0.6\textwidth, natwidth=560, natheight=350]{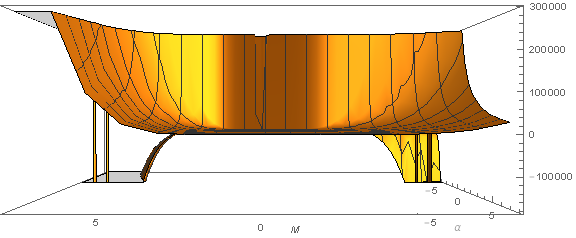}
	\caption{The same Fig. (\ref{fig:sila222}) but observed from behind. Here the dependence with respect to $T_0'(r,t)$ can be appreciated.}
	\label{fig:sila2222}
\end{figure}

\begin{figure}
	\centering
		\includegraphics[width=0.6\textwidth, natwidth=560, natheight=350]{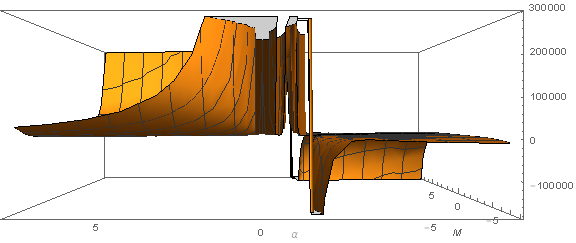}
	\caption{The same Fig. (\ref{fig:sila222}) but observed from the left. Here the dependence with respect to $\alpha$ can be appreciated.}
	\label{fig:sila22222}
\end{figure}

\begin{figure}
	\centering
		\includegraphics[width=0.6\textwidth, natwidth=560, natheight=350]{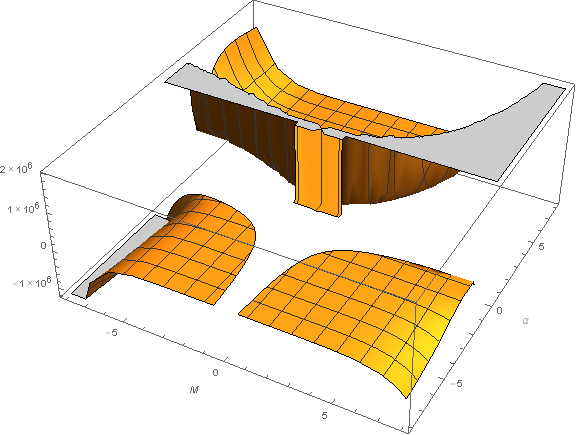}
	\caption{The square of the effective mass for the $r-r$ component. Here we can observe that the effective mass for this mode diverges for $\alpha\to0$ independent of the values taken by $T_0'(r,t)$ taken as a parameter.}
	\label{fig:sila3}
\end{figure} 

\begin{figure}
	\centering
		\includegraphics[width=0.6\textwidth, natwidth=560, natheight=350]{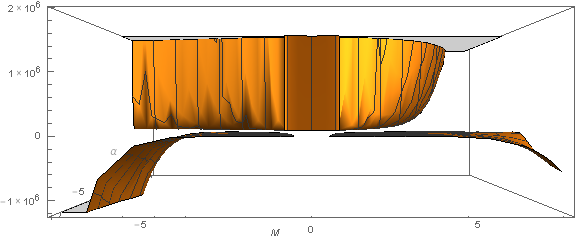}
	\caption{Fig. (\ref{fig:sila3}) but observed from behind. The plot shows a discontinuity when $T_0'(r,t)\to0$ and negative values of $\alpha$.}
	\label{fig:sila33la}
\end{figure}

\begin{figure}
	\centering
		\includegraphics[width=0.6\textwidth, natwidth=560, natheight=350]{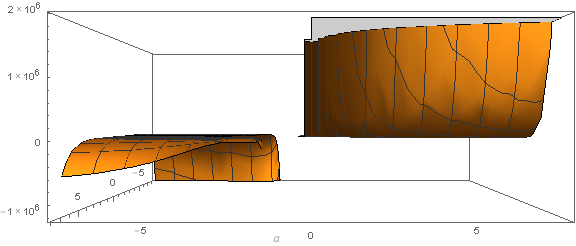}
	\caption{Fig. (\ref{fig:sila3}) but observed from the left side. Here the dependence with respect to $\alpha$ can be appreciated. again the divergence when $\alpha\to0$ appears for this case.}
	\label{fig:sila33la5}
\end{figure}
The figures (\ref{fig:sila}), (\ref{fig:sila222}) and (\ref{fig:sila3}) represent the behavior of the masses corresponding to different modes. It can be observed that the scalar and radial components of the effective mass, have similar behavior with respect to the parameter $\alpha$ and the function $T_0'(r,t)$. From the Figs. (\ref{fig:sila22}), (\ref{fig:sila22222}) and (\ref{fig:sila33la5}), it is observed that the effective masses for the different modes become very large if the parameter $\alpha$ becomes small, independent on the value taken by the arbitrary St\"uckelberg function. These results are consistent with those obtained in Fig. (\ref{fig:momokola33335577788}), where it was defined an effective graviton mass, representative for all the modes. In such a case however, the role of $T_0(r,t)$ was not clear because the analysis only covered the behavior of the parameter $\alpha$. The results are also consistent with the definition of parameter-dependent Vainshtein scale and as a consequence with the Fig. (\ref{fig:momoko3}). In other words, a large effective graviton mass, implies the no propagation of the corresponding massive modes if we select as an invariant the Vainshtein scale. On the other hand, if what we select as an invariant is the graviton mass modes, then an infinite effective Vainshtein scale is also equivalent to the no propagation of massive modes. In the case under study, it is lucky that all the modes have a similar behavior for some set of parameters. However, in general different modes have different effective masses depending on the selected values for $\alpha$ and $T_0(r,t)$ if we keep the Vainshtein scale invariant. 
	
\subsubsection{Family 2: Two free-parameters and the St\"uckelberg function well defined. The equivalence principle recovered}

\begin{figure}
	\centering
		\includegraphics[width=0.6\textwidth, natwidth=560, natheight=350]{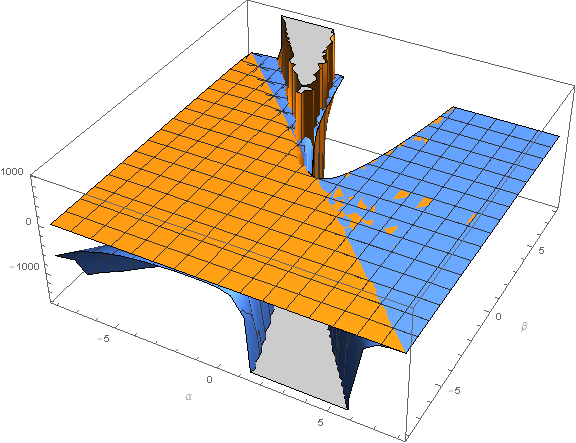}
	\caption{The square of the parameter-dependent graviton mass for the scalar mode and the radial mode. The yellow color corresponds to the positive branch of the scale parameter $S$ and the blue color corresponds to the negative branch.}
	\label{fig:sila33la55}
\end{figure}

\begin{figure}
	\centering
		\includegraphics[width=0.6\textwidth, natwidth=560, natheight=350]{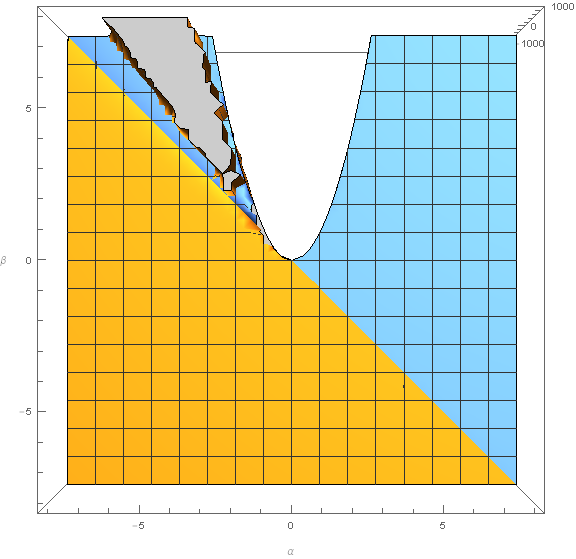}
	\caption{The same Fig. (\ref{fig:sila33la55}) but observed from above. The dependence with respect to the two free-parameters is similar with respect to the case analyzed in Sec. (\ref{eq.Ivan}) as can be perceived in Fig. (\ref{fig:momokola33335577788Alejolalamiau}).}
	\label{fig:Ivanthebest}
\end{figure}

\begin{figure}
	\centering
		\includegraphics[width=0.6\textwidth, natwidth=560, natheight=350]{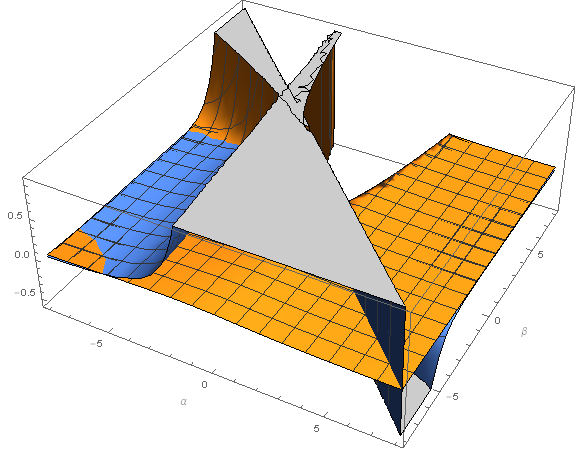}
	\caption{The square of the mass for the vectorial component. The yellow color plot corresponds to the positive branch of the scale parameter $S$ and the blue color corresponds to the negative one.}
	\label{fig:sila33la55555}
\end{figure}

\begin{figure}
	\centering
		\includegraphics[width=0.5\textwidth, natwidth=560, natheight=350]{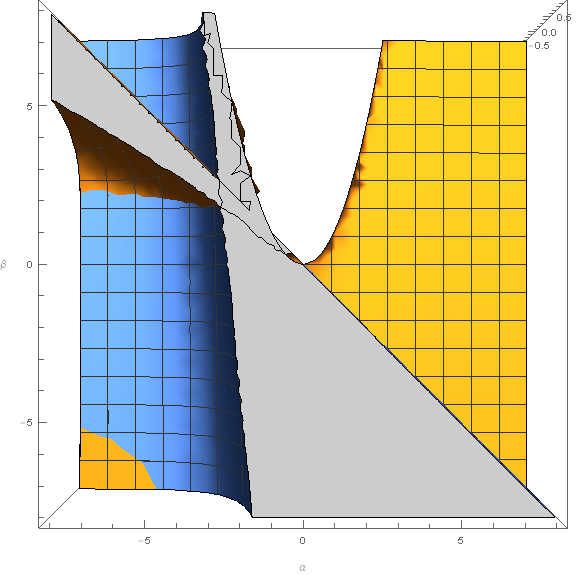}
	\caption{The same figure (\ref{fig:sila33la55555}) but observed from above. Here the dependence with respect to both, $\alpha$ and $\beta$ can be appreciated. The figure is qualitatively similar to the scalar and radial mode case analyzed in Fig. (\ref{fig:Ivanthebest}).}
	\label{fig:Ivanthebest2}
\end{figure}

As has been mentioned before, for the set of solutions with two free-parameters and $T_0(r,t)$ well defined, the equivalence principle is recovered. The potential expanded up to second order in a free-falling frame of reference is given in this case by:

\begin{eqnarray}   \label{eq:ohyeah}
\sqrt{-g}U(g,\phi)\approx S^4U(g,\phi)_{back}+h\left(\frac{S^4}{2}U(g,\phi)_{back}+S^4F(\alpha,\beta)\right)-\frac{S^4}{4}h^\alpha_{\;\;\beta}h^\beta_{\;\;\alpha}U(g,\phi)_{back} \nonumber\\
+h^2\left(\frac{S^4}{8}U(g,\phi)_{back}+\frac{S^4}{2}F(\alpha,\beta)\right).
\end{eqnarray}     
If we repeat the same procedures of the previous case, then we find that the matrix mass is given by:

\begin{equation}   \label{eq:socutethatgirl}
m^{\mu\nu}m^{\alpha\beta}=-\frac{S^4}{2}U_{back}(g,\phi)\bar{\eta}^{\mu \alpha}\bar{\eta}^{\nu \beta}+S^4\bar{\eta}^{\mu \nu}\bar{\eta}^{\alpha\beta}\left(\frac{1}{4}U(g,\phi)_{back}+F(\alpha,\beta)\right).
\end{equation}
Here we can still use the same expansion given in eq. (\ref{eq:acalanojosila}), but taking into account that the background metric now is different. In fact, for a free-falling frame of reference, the dynamical metric is simplified to be:

\begin{equation}   \label{eq:177miau}
ds^2=S^2ds^2_{M},
\end{equation} 
which is conformal to Minkowski. Then in this case, the definition of the mass for the different modes is trivial and given by:

\begin{equation}   \label{eq:177997}
m_{\mu\nu}=S^4\eta_{\mu\gamma}\eta_{\nu\beta}m^{\gamma\beta},
\end{equation}
with $\eta_{\mu\nu}$ given by the the standard Minkowski metric. From eq. (\ref{eq:socutethatgirl}), we can find the following results:

\begin{equation}   \label{eq:acala}   
(m^{00})^2=(m^{rr})^2=-\frac{1}{4}U(g,\phi)_{back}+F(\alpha,\beta),
\end{equation}

\begin{equation}   \label{eq:acala2}
(m^{0r})^2=\frac{1}{2}U(g,\phi)_{back}.
\end{equation}
The masses for the different modes can be obtained from eq. (\ref{eq:177997}) by using the results (\ref{eq:acala}) and (\ref{eq:acala2}). Figs. (\ref{fig:sila33la55}) and (\ref{fig:sila33la55555}), illustrate the behavior for the masses of the different scalar, radial and vectorial modes. Note that the scalar and radial modes have exactly the same behavior in agreement with eq. (\ref{eq:177997}). If we analyze the plane $\alpha$-$\beta$, then we can find similar behavior with respect to the case analyzed in Sec. (\ref{eq.Ivan}) in Fig. (\ref{fig:momokola33335577788Alejolalamiau}) as can be perceived the graphics (\ref{fig:Ivanthebest}) and (\ref{fig:Ivanthebest2}). However, as can be expected the behavior of the effective mass for the different modes differ. 

\begin{figure}
	\centering
		\includegraphics[width=0.6\textwidth, natwidth=560, natheight=350]{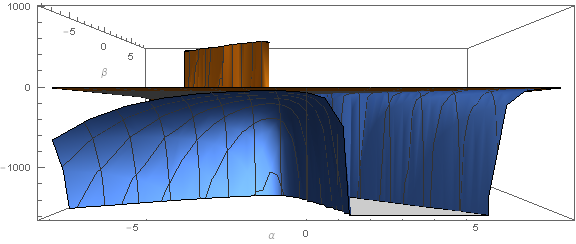}
	\caption{The same Fig. (\ref{fig:sila33la55}) but observed from the frontal perspective. Here the variation with respect to $\alpha$ can be appreciated.}
	\label{fig:sila33la555}
\end{figure}

\begin{figure}
	\centering
		\includegraphics[width=0.6\textwidth, natwidth=560, natheight=350]{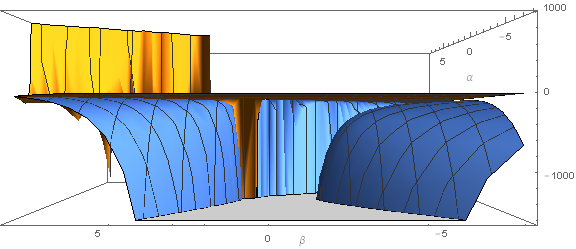}
	\caption{The same Fig. (\ref{fig:sila33la55}) but observed from the left. Here the variation with respect to $\beta$ can be appreciated.}
	\label{fig:sila33la5555}
\end{figure}

\begin{figure}
	\centering
		\includegraphics[width=0.6\textwidth, natwidth=560, natheight=350]{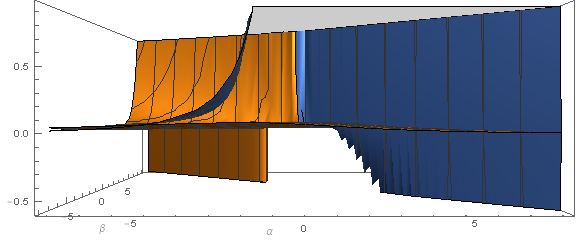}
	\caption{The same figure (\ref{fig:sila33la55555}) but observed from the frontal perspective. Here the variation with respect to $\alpha$ can be appreciated.}
	\label{fig:sila33la555555}
\end{figure}

\begin{figure}
	\centering
		\includegraphics[width=0.6\textwidth, natwidth=560, natheight=350]{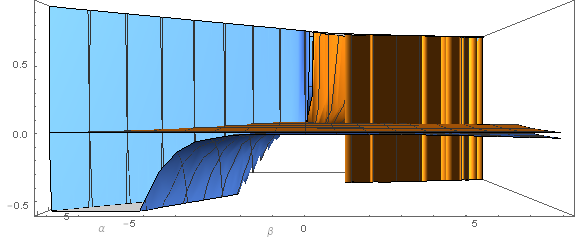}
	\caption{The same figure (\ref{fig:sila33la55555}) but observed from the right. Here the variation with respect to $\beta$ can be appreciated.}
	\label{fig:sila33la5555555}
\end{figure}

The behavior for each of the modes with respect to the parameter $\alpha$ or $\beta$, reveal the existence of some peaks as can be seen from Figs. (\ref{fig:sila33la555}) and (\ref{fig:sila33la5555}) for the scalar and radial modes and from Figs. (\ref{fig:sila33la555555}) and (\ref{fig:sila33la5555555}) for the vectorial one. For the scalar and vectorial modes, the observed peaks are very large. The peaks in such a case, can be interpreted as the combination of parameters for which the mode becomes heavy and does not propagate so far. This is equivalent to a screening mechanism effect. In addition, we can observe that the peaks for the vectorial component are not so large. This means that the vectorial modes in this case can propagate freely for most of the parameter combinations of the theory. The vectorial modes however, normally are decoupled from matter when we analyze the theory at the decoupling limit level \cite{derham}. \\\\\\\\\\\\\\\\\\\\

\section{Summary}
Here I have derived the simplest method in order to predict whether the Vainshtein screening mechanism might appear for some specific parameter combinations of the theory. The method is based on the Vainshtein conditions derived by the author in \cite{My papers}. The details of the potential are not important for the present analysis. The regions where the Vainshtein mechanism operates are not necessarily relevant from the the physical point of view. This means that the conditions derived in this manuscript are necessary but not sufficient in order to guarantee the recovery of GR. The recovery of GR not only implies the predictions at the solar system scale but also the predictions related to the physics of black-holes. In \cite{On the app} for example, it was discovered that some parameter combinations can reproduce branch point effects when we analyze the periodicity structure of the propagators.  
This manuscript is developed by using two equivalent points of view. The first one is worked by fixing the graviton mass and then deriving an effective Vainshtein scale. The second point of view is developed by fixing the Vainshtein scale and then deriving an effective version of the graviton mass. Here I consider that both points of view are physically equivalent and considering one or the other is a matter of choice. Finally in order to perform a rigorous analysis, I have studied the behavior of the effective mass for each of the modes, namely, scalar, vectorial and radial. In this case I have also maintained the effective Vainshtein scale as an invariant. The analysis revealed that the mass for the different modes can behave in a different way depending on the parameter combinations considered. In other words, although from the qualitative point of view some cases look similar, the detailed study revealed that the effective mass for the different modes are not the same in general. In particular the scalar and radial modes have the same effective mass for the case of two free-parameters and $T_0(r,t)$ constrained. In such a case, only the vectorial component behaves different. However, if we look at the case with one free-parameter and $T_0(r,t)$ arbitrary, then the different modes behave different when we look at them in detail. However, even in this situation, the scalar and radial components have similar behavior. The difference comes from the behavior with respect to the arbitrary St\"uckelberg function. In fact, this function defines the direction of the preferred notion of time. It would be interesting to extend these ideas to other solutions like the ones proposed in \cite{Sari} among others, where other gravitational contributions might appear. The method could also be extended for the analysis of more complicated solutions like the Kerr one, discovered recently inside the massive gravity formulations \cite{Babi}. For the case of solutions containing the time variable inside the metric components, the Vainshtein conditions (\ref{eq:ptm}) cannot ignore the time variation in such a case. Then the parameter combinations where the Vainshtein (screening) mechanism appears might change with time. \\\\

{\bf Acknowledgement}\\
I.A is supported by the CAS PIFI program for Post-Doctoral researchers.

\end{document}